\documentclass[twocolumn,showpacs,preprintnumbers,amsmath,amssymb]{revtex4}

\usepackage{graphicx}
\usepackage{bm}

\usepackage{amsmath,amscd}
\usepackage{amssymb}
\usepackage[dvips]{color}

\begin{document}
\definecolor{gris}{rgb}{.52,.52,.52}

\title{From DNA sequence analysis to modelling replication in the human genome}

\author{E.B. Brodie of Brodie$^1$}
\author{S. Nicolay$^1$}
\author{M. Touchon$^2$}
\author{B. Audit$^1$}
\author{Y. d'Aubenton-Carafa$^2$}
\author{C. Thermes$^2$}
\author{A. Arneodo$^1$}
\affiliation{$^1$ Laboratoire Joliot Curie, Ecole Normale Sup\'erieure de Lyon, 46 All\'ee d'Italie,
69364 Lyon Cedex 07, France\\
$^2$ Centre de G\'en\'etique Mol\'eculaire (CNRS), All\'ee de la Terrasse,
91198 Gif-sur-Yvette, France}

\date{\today}

\begin{abstract}
  We explore the large-scale behavior of nucleotide compositional
  strand asymmetries along human chromosomes.
  As we observe for 7 of 9 origins
  of replication experimentally identified so far,
  the $(TA+GC)$ skew displays rather sharp upward jumps,
  with a linear decreasing profile in between
  two successive jumps. We present a model
  of replication with well positioned replication origins and random
  terminations that accounts for the observed characteristic serrated
  skew profiles. We succeed in identifying 287 pairs of
  putative adjacent replication origins with
  an origin spacing $\sim 1$--$2 Mbp$, that are likely to
  correspond to replication foci
  observed in interphase nuclei and recognized as stable structures
  that persist throughout subsequent cell generations.
\end{abstract}

\pacs{87.15.Cc, 87.16.Sr, 87.15.Aa}
\maketitle
\begin{figure}
  \includegraphics*[scale=0.45]{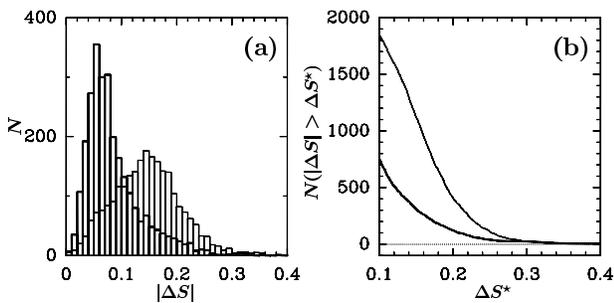}%
\caption{\label{histosaut} Statistical analysis of the sharp
 jumps detected in the $S$ profiles
 of the 22 human autosomal chromosomes by the WT microscope
 at scale $a=200~kbp$ for repeat--masked
 sequences ~[12(b)]. $|\Delta S| = |\overline{S}(3') - \overline{S}(5')|$, where
 the averages were computed over the two adjacent $20~kbp$ windows
 respectively in the 3' and 5' direction from the detected jump location.
 (a) Histograms $N(|\Delta S|)$ of $|\Delta S|$ values.
 (b) $N(|\Delta S|>\Delta S^\ast)$ vs $\Delta S^\ast$.
 In (a) and (b), the solid (resp. thin) line corresponds to downward $\Delta S < 0$
 (resp. upward $\Delta S > 0$) jumps.\vspace{-0.9cm}\\}
\end{figure}
\begin{figure}
   \includegraphics*[scale=0.38]{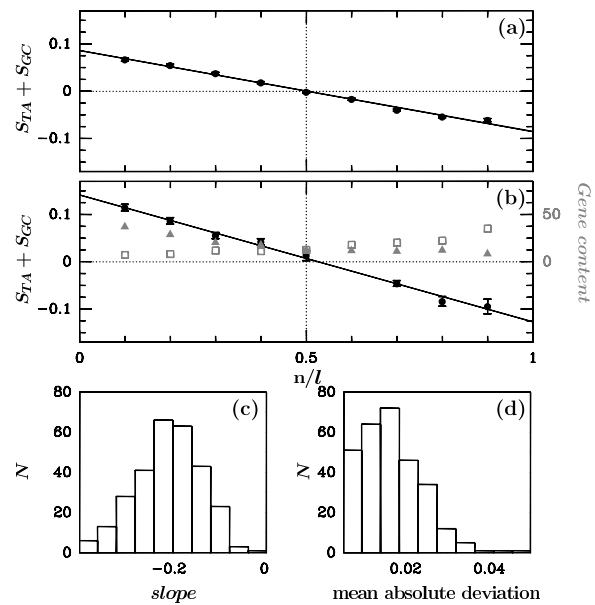}
   \caption{\label{replicon} Statistical analysis of the skew
     profiles of the 287 pairs of \emph{ori} selected as explained in the text.
     The \emph{ori} spacing
     $l$ was rescaled to 1 prior to computing the mean $S$ values
     in windows of width $1/10$, excluding from the analysis the first and
     last half intervals. (a) Mean $S$ profile ($\bullet$) over windows that
     are more than 90\% intergenic. (b) Mean $S$ profile ($\bullet$) over
     windows that are more than 90\% genic; the symbols (\textcolor{gris}{$\blacktriangle$})
     (resp. (\textcolor{gris}{$\square$})) correspond to the percentage of sense (antisense)
     genes located at that position among the 287 putative \emph{ori} pairs. (c) Histogram
     of the slope $s$ of the skew profiles after
     rescaling $l$ to 1. (d) Histogram of the mean
     absolute deviation of the $S$ profiles from a linear profile.
     \vspace{-0.9cm}\\
   }
 \end{figure}
\begin{figure}
  \includegraphics*[scale=0.43]{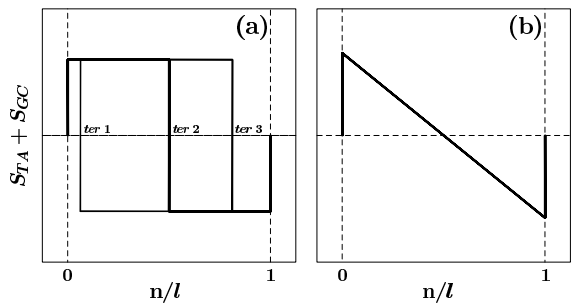}%
\caption{\label{model} A model for replication in the human genome.
 (a) Theoretical skew profiles obtained when assuming
 that two equally active adjacent \emph{ori}
 are located at $n/l = 0$ and $1$, where $l$ is the
 \emph{ori} spacing; the 3 profiles in
 thin, thick and normal lines, correspond to different
 \emph{ter} positions. (b) Theoretical mean $S$ profile
 obtained by summing step--like profiles as in (a), under
 the assumption of a uniform random positioning of the \emph{ter}
 in between the two \emph{ori}.
 \vspace{-0.9cm}\\
}
\end{figure}

DNA replication is an essential genomic function responsible for
the accurate transmission of genetic
information through successive cell generations. According to the
``replicon'' paradigm derived from prokaryotes \cite{un},
this process starts with the binding of some
``initiator'' protein to a specific ``replicator'' DNA sequence
called origin of replication (\emph{ori}).
The recruitement of additional factors initiates the bidirectional
progression of two divergent replication forks along the chromosome.
One strand is replicated continuously from the origin (leading strand),
while the other strand is replicated in discrete steps towards the origin
(lagging strand).
 In eukaryotic cells, this event is initiated
 at a number of \emph{ori} 
 and propagates until two converging forks
 collide at a terminus of replication (\emph{ter})
 \cite{trois}. The initiation
 of different \emph{ori} is coupled to the cell cycle
 but there is a definite flexibility in the usage of the \emph{ori}
 at different developmental stages \cite{quatre,six}.
Also, it can be strongly influenced by the distance and timing
of activation of neighbouring \emph{ori}, by the transcriptional
activity and by the local chromatin structure \cite{quatre}.
Actually, sequence requirements for an \emph{ori} vary significantly
between different eukaryotic organisms. In the unicellular
eukaryote \emph{Saccharomyces cerevisiae}, the \emph{ori} spread over
100--150 $bp$ and present some highly conserved motifs
\cite{trois}. 
In the fission yeast \emph{Schizosaccharomyces pombe}, there is no
clear consensus sequence and the \emph{ori} spread over at least
800 to 1000 $bp$ \cite{trois}. In multi--cellular
organisms, the \emph{ori} are rather poorly defined
and initiation may occur at multiple sites distributed over
thousands of base pairs \cite{cinq}.
 Actually, cell diversification may have led higher eukaryotes to develop various
 epigenetic controls over the \emph{ori} selection rather than to conserve
 specific replicator sequences \cite{huit}. This might explain
 that only very few \emph{ori} have
 been identified so far in multi--cellular eukaryotes,
 namely around 20 in metazoa and only about 10 in human \cite{trentecinq}.
The aim of the present work is to show that with an appropriate
 coding and an adequate methodology, one can challenge the issue
 of detecting putative \emph{ori} directly from the
 genomic sequences.\\
\indent According to the second parity rule \cite{lobry95},
 under no--strand bias conditions, each genomic DNA strand
 should present equimolarities of A and T and of G and C.
 Deviations from intrastrand equimolarities have
 been extensively studied in prokaryotic, organelle and viral
 genomes for which they have been used to detect 
 the \emph{ori} \cite{treize}.
 Indeed the GC and TA skews abruptly switch sign at the \emph{ori}
 and \emph{ter} displaying step like profiles, such that the leading strand
 is generally richer in G than in C, and to a lesser
 extent in T than in A. During replication, mutational events
 can affect the leading and lagging strands differently, and an
 asymmetry can result if one strand incorporates more mutations
 of a particular type or if one strand is more efficiently
 repaired \cite{treize}.
 In eukaryotes, the existence of compositional biases has been debated
 and most attempts to detect the \emph{ori}
 from strand compositional asymmetry have been inconclusive.
 In primates, a comparative study of the $\beta$--globin \emph{ori} has
 failed to reveal the existence of a replication--coupled mutational
 bias \cite{dixsept}.
 Other studies have led to rather opposite results.
 The analysis of the yeast genome
 presents clear replication--coupled strand asymmetries
 in subtelomeric chromosomal regions \cite{seize}.
 A recent space--scale analysis \cite{vingtquatre} of the
 GC and TA skews in $Mbp$ long human contigs has
 revealed the existence of compositional strand asymmetries in
 intergenic regions,
 suggesting the existence of a replication bias.
 Here, we show that the $(TA+GC)$ skew profiles of the
 22 human autosomal chromosomes, display a remarkable serrated
 ``factory roof'' like behavior that differs from the crenelated
 ``castle rampart'' like profiles resulting from the prokaryotic
 replicon model \cite{treize}. This observation will lead us to propose an
 alternative model of replication in higher eukaryotes.\\
\indent Sequences and gene annotation data were downloaded 
from the UCSC Genome Bioinformatics site and correspond to the assembly
 of July 2003 of the human genome.
To exclude repetitive elements that might have been inserted
 recently and would not reflect long--term evolutionary patterns,
 we used the repeat-masked version of the genome leading to a homogeneous
 reduction of $\sim 40-50\%$ of sequence length.
 All analyses were
 carried out using ``knowngene'' gene annotations. The TA and GC skews
 were calculated as $S_{TA}=(T-A)/(T+A)$ and $S_{GC}=(G-C)/(G+C)$.
 Here, we will mainly consider $S=S_{TA}+S_{GC}$, since
 by adding the two skews, the sharp transitions of interest are significantly
 amplified.\\
\begin{figure}
  \includegraphics*[scale=.7]{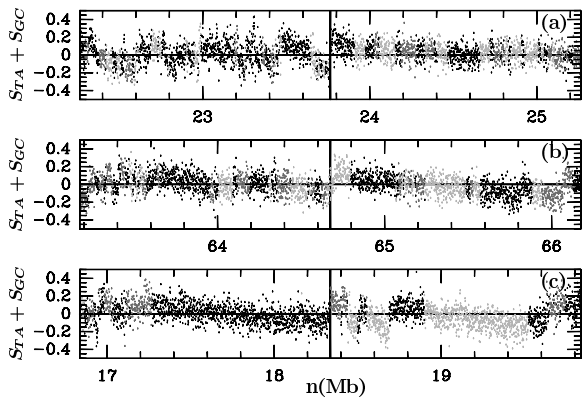}%
\caption{\label{oriconnues} $S=S_{TA}+S_{GC}$ \emph{vs} the
 position $n$ in the repeat--masked sequences, in
 regions surrounding 3 known human \emph{ori} (vertical bars):
 (a) MCM4 (native position $48.9~Mbp$ in chr. 8 [7(b)]); (b) c--myc
 (nat. pos. $128.7~Mbp$ in chr. 8 [7(a)]); (c) TOP1 (nat. pos.
 $40.3~Mbp$ in chr. 20 [7(c)]).
 The values of $S_{TA}$ and $S_{GC}$ were calculated in adjacent 1~$kbp$
 windows. The dark (light) grey dots
 refer to ``sense'' (``antisense'') genes
 with coding strand identical (opposed) to the sequence;
 black dots correspond to intergenes.\vspace{-0.9cm}\\}
\end{figure}
\indent In Fig.~\ref{oriconnues} are shown the skew $S$ profiles of 3
fragments of chromosomes 8 and 20 that contain 3
experimentally identified \emph{ori}.
As commonly observed for eubacterial genomes \cite{treize},
these 3 \emph{ori} correspond to rather sharp (over
 several $kbp$) transitions from negative to positive $S$ values that
 clearly emerge from the noisy background. The leading strand is
 relatively enriched in T over A and in G over C. The investigation
 of 6 other known human \emph{ori} \cite{trentecinq}
 confirms the above observation for at least 4 of them (the 2 exceptions,
 namely the Lamin B2 and $\beta$--globin \emph{ori}, might well be inactive
 in germline cells or less frequently used than the adjacent \emph{ori}).
 According to the gene environment, the amplitude
 of the jump can be more or less important and its position more or
 less localized (from a few $kbp$ to a few tens $kbp$).
 Indeed, it is known that transcription generates positive
 TA and GC skews on the coding strand \cite{vingttrois,vingtdeux},
 which explains that  larger jumps are observed when the sense genes
 are on the leading strand and/or
 the antisense genes on the lagging strand, so that
 replication and transcription biases add to each other.
 On the contrary to the
 replicon characteristic step like profile observed for
 eubacteria \cite{treize}, $S$ is definitely not
 constant on each side of the \emph{ori} location making quite elusive the
 detection of the \emph{ter} since no corresponding 
 downward jumps of similar amplitude can be found in Fig.~\ref{oriconnues}.\\
\begin{figure}
  \includegraphics*[scale=.7]{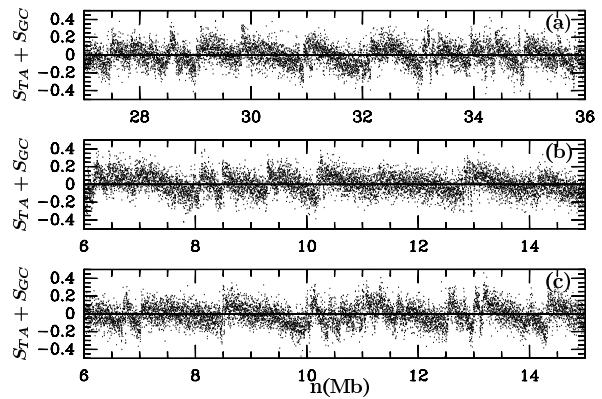}%
\caption{\label{profilbiais}$S=S_{TA}+S_{GC}$ skew profiles in 9~$Mbp$ 
  repeat-masked fragments in the human chromosomes 9 (a), 14 (b) and 21 (c). 
  Qualitatively similar but less spectacular serrated
  $S$ profiles are obtained with the native human sequences.\vspace{-0.9cm}\\}
\end{figure}
\indent In Fig.~\ref{profilbiais} are shown the $S$ profiles of
 long fragments of chromosomes 9, 14  and 21,
 that are typical of a fair proportion of the $S$ profiles
 observed for each chromosome.
 Sharp upward
 jumps of amplitude ($\Delta S \sim 0.2$) similar to the ones
 observed for the known \emph{ori} in Fig.~\ref{oriconnues},
 seem to exist also at many other locations along the human
 chromosomes. But the most striking feature is the fact that in between two
 neighboring major upward jumps, not only the noisy $S$
 profile does not present any comparable downward sharp transition,
 but it displays a remarkable decreasing linear behavior.
At chromosome scale, one thus gets jagged $S$ profiles
 that have the aspects of ``factory roofs'' rather than
 ``castle rampart'' step like profiles as expected for the
 prokaryotic replicon model \cite{treize}.
The $S$ profiles in Fig.~\ref{profilbiais} look somehow
 disordered because of the extreme variability in the distance
 between two successive upward jumps, from
 spacings $\sim$ 50--$100~kbp$ ($\sim100$--$200~kbp$
 for the native sequences) up to 2--3$~Mbp$
 ($\sim4$--$5~Mbp$ for the native sequences) in agreement with
 recent experimental studies that have shown that mammalian replicons
 are heterogeneous in size with an average size $\sim 500~kbp$,
 the largest ones being as large as a few $Mbp$ \cite{trentedeux}.
We report in
 Fig.~\ref{histosaut} the results of a systematic detection of upward and
 downward jumps using the wavelet--transform (WT) based methodology described
 in Ref. [12(b)]. The selection criterium was to
 retain only the jumps corresponding to discontinuities in the $S$ profile
 that can still be detected with the WT
 microscope up to the scale $200~kbp$ which is smaller than the typical
 replicon size and larger than the typical gene size.
 In this way, we reduce the contribution of jumps associated with
 transcription only and maintain a good sensitivity to replication induced jumps.
 A set of 5100 jumps was detected
 (with as generally expected an almost equal proportion of upward and downward jumps).
 In Fig.~\ref{histosaut}(a) are reported the histograms of the amplitude
 $|\Delta S|$ of the so--identified upward ($\Delta S > 0$) and downward
 ($\Delta S < 0$) jumps respectively, for the repeat--masked sequences.
 These histograms do not superimpose, the former being
 significantly shifted to larger $|\Delta S|$ values.
 When plotting
 $N(|\Delta S|>\Delta S^\ast)$ vs $\Delta S^\ast$ in Fig.~\ref{histosaut}(b),
 one can see that the number of large amplitude upward jumps overexceeds
 the number of large amplitude downward jumps.
These results confirm that most of the sharp upward transitions in the $S$
 profiles in Figs \ref{oriconnues} and \ref{profilbiais}, have no sharp
 downward transition counterpart. This demonstrates that these jagged
 $S$ profiles are likely to be representative of a general asymmetry
 in the skew profile behavior along the human chromosomes.\\
\indent As reported in a previous work \cite{vingtdeux}, the analysis of a complete
set of human genes revealed that most of them present TA and GC skews and
 that these biases are correlated to each other and are specific to
 gene sequences. One can thus wonder to which extent
 the transcription machinery can account for the jagged $S$ profiles
 shown in Figs \ref{oriconnues} and \ref{profilbiais}. According to the
 estimates obtained in Ref. \cite{vingtdeux}, the mean jump amplitudes
 observed at the transition between transcribed and
 non--transcribed regions are $|\Delta S_{TA}|\sim 0.05$ and
 $|\Delta S_{GC}|\sim 0.03$ respectively.
The characteristic amplitude of a transcription induced transition
 $|\Delta S| \sim 0.08$ is thus
 significantly smaller than the amplitude $\Delta S \sim 0.20$
 of the main upward jumps in Fig.~\ref{profilbiais}.
 Hence, it is possible that, at the transition between an
 antisense gene and a sense gene,
 the overall jump from negative to positive $S$ values may reach
 sizes $\Delta S \sim 0.16$ that can be comparable to the ones of the upward
 jumps in Fig.~\ref{profilbiais}.
 However, if some co--orientation of the transcription and replication
 processes may account for some of the sharp upward transitions in
 the skew profiles,
 the systematic observation
 of ``factory roof'' skew scenery in intergenic regions
 as well as in transcribed regions, strongly suggests that
 this peculiar strand bias is likely to originate from the replication
 machinery.
To further examine if intergenic regions present typical ``factory roof''
skew profiles, we report in Fig.~\ref{replicon} the results of the statistical
 analysis of
 287 pairs of putative adjacent \emph{ori} that actually
 correspond to 486 putative \emph{ori} almost equally distributed among the
 22 autosomal chromosomes. These putative \emph{ori} were identified by (i)
 selecting pairs of successive jumps of amplitude $\Delta S \geq 0.12$,
 and (ii) checking that none of these upward jumps could be explained by an antisense
 gene --- sense gene transition.
 In Fig.~\ref{replicon}(a) is shown the $S$ profile obtained after
 rescaling the putative \emph{ori} spacing $l$ to 1 prior to computing the average
 $S$ values in windows of width $1/10$ that contain more than 90\%
 of intergenic sequences. This average profile is linear and crosses
 zero at the median position $n/l=1/2$, with an overall upward jump
 $\Delta S \simeq 0.17$. The corresponding average $S$ profile
 over windows that are now more than 90\% genic is shown in
 Fig.~\ref{replicon}(b). A similar linear profile is obtained
 but with a jump of larger mean amplitude $\Delta S \simeq 0.28$.
 This is a direct consequence
 of the gene content of the selected regions.
 As shown in Fig.~\ref{replicon}(b),
 sense (antisense) genes are preferentially on the left (right) side
 of the 287 selected sequences, which implies that the replication
 and -- when present -- transcription biases tend to add up.
 In Fig.~\ref{replicon}(c) is shown
 the histogram of the linear slope values
 of the 287 selected skew profiles after rescaling their length to 1.
 The histogram of mean absolute deviation
 from a linear decreasing profile reported in
 Fig.~\ref{replicon}(d), confirms the linearity of each selected skew
 profiles.\\
 \indent 
 Following these observations, we propose in Fig.~\ref{model} a rather crude model for
 replication that relies on the hypothesis that the \emph{ori} are
 quite well positioned while the \emph{ter} are randomly
 distributed. In other words, replication would proceed in a
 bi--directional manner from well defined initiation positions,
 whereas the termination would occur at different positions from
 cell cycle to cell cycle \cite{trentequatre}. Then if one assumes
 that (i) the \emph{ori} are identically active and (ii)
 any location in between two adjacent \emph{ori} has an equal probability
 of being a \emph{ter}, the continuous superposition of step--like profiles
 like those in Fig.~\ref{model}(a) leads to the anti--symmetric 
 skew pattern shown in Fig.~\ref{model}(b), \emph{i.e.} a
 linear decreasing $S$ profile that
 crosses zero at middle distance from the two \emph{ori}.
 This model is in good agreement with the overall properties of the
 skew profiles observed in the human genome and sustains the hypothesis
 that each detected upward jump corresponds to an \emph{ori}.\\
\indent To summarize, we have proposed a simple model for replication in the
human genome whose key features are (i) well positioned \emph{ori} and
(ii) a stochastic positioning of the \emph{ter}.
This model predicts jagged skew profiles as observed around
most of the experimentally identified \emph{ori} as well
as along the 22 human autosomal chromosomes.
Using this model as a guide, we have selected 287 domains
delimited by pairs of successive upward jumps in the $S$ 
profile and covering 24\% of the genome. 
The 486 corresponding jumps are likely to mark 486 \emph{ori}
active in the germ line cells.
As regards to the rather large size of the selected sequences
($\sim 2~Mbp$ on the native sequence),
these putative \emph{ori} are likely to correspond
to the large replicons that require most of the S--phase
to be replicated \cite{trentedeux}.
Another possibility is that these \emph{ori}
might correspond to the so--called replication foci observed
in interphase nuclei \cite{trentedeux}. These stable
structures persist throughout the cell cycle and
subsequent cell generations, and likely represent
a fundamental unit of chromatin organization.
Although the prediction of 486 \emph{ori} seems a significant
achievement as regards to the very small number of experimentally
identified \emph{ori},
one can reasonably hope to do much better relatively to the
large number (probably several tens of thousands) of
\emph{ori}.
Actually what
makes the analysis quite difficult is the extreme variability
of the \emph{ori} spacing from $100~kbp$ to several $Mbp$,
together with the necessity of
disentangling the part of the strand asymmetry coming from
replication from that induced by transcription, a task
which is rather delicate in regions with high gene density.
To overcome these difficulties, we plan to use the WT with
the theoretical skew profile in Fig.~\ref{model}(b)
as an adapted analyzing wavelet.
The identification of a few thousand putative \emph{ori} in the human genome
would be a very promising methodological step
towards the study of replication in mammalian genomes.\\
\indent This work was supported by the Action Concert\'ee Incitative IMPbio 2004, the
Centre National de la Recherche Scientifique, the French Minist\`ere de la Recherche
and the R\'egion Rh\^one-Alpes.

\end{document}